\documentclass[runningheads]{llncs}
\usepackage{graphicx}
\usepackage{comment}
\usepackage{xcolor}
\usepackage{hyperref}
\usepackage{svg}
\usepackage{amsmath}
\usepackage{caption}
\usepackage{subcaption}
\usepackage{listings}
\usepackage{color}
\usepackage[T1]{fontenc}

\usepackage{algorithmic}
\usepackage{algorithm}
\usepackage{multirow}

\usepackage{soul}
\definecolor{aliceblue}{rgb}{0.94, 0.97, 1.0}

\PassOptionsToPackage{hyphens}{url}\usepackage{hyperref}
\usepackage{xspace}
\usepackage{enumitem}

\usepackage{amsmath}
\usepackage{comment}
\usepackage{fancybox}

\newcounter{rqcounter}

\newcounter{defn}

\newcounter{observation}

\newcommand{\Case}[1]{#1}

\newcommand{\RFix}[1]{}

\newcommand{\Space}[1]{}

\newcommand{\testcase}{test\Case{ case}}

\newcommand{\Deltadebugging}{Delta-debugging}
\newcommand{\deltadebugging}{delta-debugging}

\newcommand{\CCoverage}{$C\%$-Coverage}

\newcommand{\OT}{\ensuremath{t_o}}
\newcommand{\RT}{\ensuremath{t_r}}
\newcommand{\Cov}[1]{\ensuremath{Cov(#1)}}

\newcommand{\Size}[1]{\ensuremath{Size(#1)}}

\newcommand{\approach}{\textsc{Diar}\xspace}
\begin{document}

\title{\approach: Removing Uninteresting Bytes from Seeds in Software Fuzzing}

%
%
%
%
%

\author{
    Aftab Hussain, 
    Mohammad Amin Alipour
}
\institute{
    \email{ahussain27@uh.edu, maalipou@central.uh.edu }\\
    University of Houston, Houston, TX, USA
}

\maketitle              
\begin{abstract}
Software fuzzing mutates bytes in the test seeds to explore different behaviors of the program under test. 
Initial seeds can have great impact on the performance of a fuzzing campaign. Mutating a lot of uninteresting bytes in a large seed wastes the fuzzing resources. 
In this paper, we present the preliminary results of our approach that aims to improve the performance of fuzzers through identifying and removing uninteresting bytes in the seeds. In particular, we present \approach, a technique that reduces the size of the seeds based on their coverage. 
Our preliminary results suggest fuzzing campaigns that start with reduced seeds, find new paths faster, and can produce higher coverage overall.

\end{abstract}

\section{Introduction}
\label{intro}

Coverage-guided fuzzers, fuzzers for short, have become an important tool in testing
software systems and uncovering bugs and software vulnerabilities. 
Due to their easy-to-use design and proven potential of finding bugs and vulnerabilities, they are increasingly being adopted in the industry, and popular coverage-guided fuzzers, like AFL~\cite{afl} and AFL++~\cite{aflpp}, are regularly used  for testing applications at large companies. 

These fuzzers mutate bytes in the seeds to generate new tests, and coverage feedback steers this test generation. There is a large body of work that has been dedicated to improve the performance of the fuzzers. The majority of work in this area has been concerned with the mutation operators and scheduling in steering the test generation. However, seed selection has just recently gained some interest~\cite{seedselection:2021:issta} and most papers treat that ``casually''~\cite{Klees:EvaluatingFuzzTesting:2008}.

Seeds can impact the performance of fuzzing. While a small seed filled with bytes associated with interesting behavior can help fuzzers explore a larger state space of programs faster, a large seed with many uninteresting bytes, e.g., payload data in network protocols, that do not contribute to the interesting behaviors of the program, will trap the fuzzers in a long sequence of futile mutations of bytes. As one solution to this program could be to remove such large seeds, we note that in some settings the seeds can be inherently large, e.g., object or media files -- hence such solutions are inapplicable. Therefore, we need approaches to preprocess the seeds to identify  uninteresting bytes and somehow exclude them from mutation.

In this paper, we introduce \approach, an approach that aims to identify and remove uninteresting bytes from the seeds. \approach uses non-adequate test reduction~\cite{non-adequate-reduction} to remove chunks of a seed that do not contribute much to the test coverage of the seed.
\approach is customizable and allows users to identify a time budget for the reduction and preset the maximum allowed loss of test coverage.
We performed a preliminary study of the approach on two subjects: \texttt{xmllint} and \texttt{readelf}, programs from widely used libraries libxml2 (an XML parser) and binutils (a collection of GNU binary tools).
We found that in the case of the input that contains a substantial number of uninteresting bytes, i.e., \texttt{readelf}'s input, \approach can substantially reduce the seed, and fuzzing campaigns that use the reduced test, cover a larger portion of the program under test and find more crashes.
Our initial results are encouraging and provide direction for future research. 

\noindent\textbf{Contributions.} The main contributions of this paper are as follows:
 \begin{itemize}
    \item We propose \approach\footnote{We look forward to making DIAR's script available at the time of publication.}, a novel customizable approach, based on non-adequate test reduction for removing uninteresting bytes in large seeds.
    \item We present the preliminary results of applying this technique in fuzzing two programs with AFL. 
\end{itemize}

\noindent\textbf{Paper Organization.} We lay out the foundations of \approach in Section~\ref{method}. In Sections~\ref{sec:settings} and~\ref{sec:results}, we detail the design of our experiments and present our results. In Section~\ref{sec-disc}, we discuss the implications of this work and provide future directions. We present some related literature in Section~\ref{rel-work}. Finally, we conclude our paper in Section~\ref{concl}.

\section{\approach}
\label{method}

In this section, we first describe the workflow of \approach (\ref{subsec-workflow}). Then we explain \approach's reduction criteria and the rationale for choosing them (\ref{subsec-cons}).

\subsection{\approach Workflow}
\label{subsec-workflow}

\approach takes as input a target binary of a test subject program along with a valid test seed that we wish to reduce. 
At first the delta debugging module will reduce the seed based on the delta-debugging algorithm outlined in~\cite{zeller-tse-2002} to produce a  1-minimal seed that satisfies a set of constraints that we defined. These constraints include coverage similarity, reduction ratio, and execution status. We explain these constraints in detail in the subsequent subsection. For brevity, we refer the readers to~\cite{zeller-tse-2002} for details of the delta-debugging algorithm.

\subsection{Constraints}
\label{subsec-cons}

\noindent\textit{Coverage Similarity.} We use the \CCoverage{} reduction constraint introduced in~\cite{non-adequate-reduction}\Space{ ($cov\_similarity$ in Algorithm~\ref{alg:ddmin})}. It specifies that a reduced \testcase{} \RT{} must cover at least $C\%$ of the statements\Space{ that are} covered by the
    original \testcase{} \OT:

        \begin{center}
            $\frac{|\Cov{\RT}\cap\Cov{\OT}|}{|\Cov{\OT}|}\ge C\%$
        \end{center}
Since the above constraint does not require preserving all code covered by the original \testcase{}, as in the stricter version in~\cite{icst14,STVR:STVR1574}, a reduced \testcase{} that satisfies the above constraint is defined as a \textit{non-adequately} reduced seed~\cite{non-adequate-reduction}. 

It has been shown that \textit{non-adequately} reduced seeds can still be useful in software testing~\cite{non-adequate-reduction}, a finding we also corroborate from our results in the fuzzing realm. Moreover, non-adequate reduction is faster to compute~\cite{non-adequate-reduction}; 
our experiments show that adequate test reduction that preserves exact coverage of the original seeds can be very expensive. 



\noindent\textit{Reduction Ratio.} We found that even non-adequate reduction using \deltadebugging{} can yield reduced seeds that are almost similar to the original seed (e.g., for some seeds of a binary of libxml2, the percentage reduction in size was less than 5\%). To counter this systematically, we introduce the notion of $R\%$-reduction and augment the \deltadebugging{} algorithm with this new constraint. According to the constraint, a reduced seed \RT{} is accepted only if it has been obtained by reducing the size of the original seed \OT~by at least $R\%$:
        \begin{center}
            $\frac{Size(\OT)-Size(\RT)}{Size(\OT)}\ge R\%$
        \end{center}

\noindent\textit{Exit Status.} 
Reduction may produce tests with different behavior, and can sometimes even find crashes. However, for consistent experimentation, we require the exit code of the execution of the reduced tests on the program be the same as that of the original test. 
Therefore, \approach captures the status of the seed's execution on exit, and introduces the exit status constraint in the \deltadebugging{} algorithm. This constraint requires the exit statuses of \OT{} and \RT{} to be equal. 

\section{Experimental Settings}
\label{sec:settings}

In this section, we outline the overall design of our preliminary experiments and the test subjects we used. 


We evaluated two test subjects in our experiments: \texttt{xmllint} (from libxml2 library), and \texttt{readelf} (from binutils library). 
In the first phase of our experiments, we used \approach to obtain a reduced test case from an original test case for each subject.
In the second phase, we fuzzed each test subject with AFL fuzzer. In total, we performed six 
24-hour fuzzing jobs for each test subject; three AFL fuzzing instances using the original 
test case as the input, and three instances using the reduced test case as the input. We thus performed a total 
of 288 CPU hours of fuzzing. Both phases were performed on a server machine with 12 Intel(R) 1.90GHz Xeon(R) CPUs and 64 GB RAM with Ubuntu 18.04.5 LTS.

\subsection{Test Subjects}

Table~\ref{tab-subs} shows the details of the target programs we evaluated 
in our experiments, including the execution command that was used for each of them (``@@" was only used during fuzzing, which is an AFL directive indicating the position in the command where the name of the input test case folder is 
placed). 

\begin{table}
\caption{Programs evaluated in our experiments.}

\setlength{\tabcolsep}{3pt}
\centering
\begin{tabular}{llll}
\hline
\textbf{Target}  & \textbf{Source} & \textbf{Input}  & \textbf{Execution}         \\ 
\textbf{program} &  \textbf{(ver./commit-id)} &  \textbf{format} &  \textbf{command}        \\ \hline
\texttt{xmllint}        & libxml2 (1fbcf40)            & text (xml)   & \texttt{xmllint -o /dev/null @@ }\\
\texttt{readelf}        & binutils (2.30)              & binary       & \texttt{readelf -a @@}           \\
\hline
\end{tabular}
\label{tab-subs}
\end{table}

\section{Results}
\label{sec:results}
In this section, we present the results of our experiments, where we investigate the following research questions:

\begin{itemize}[leftmargin=.38in]
 \item[RQ1.] How effective is \approach in reducing seeds?
 \item[RQ2.] What is the impact of reduced seeds on exploration of new paths in fuzzing?
 \item[RQ3.] Do reduced seeds lead to more crashes and coverage?

\end{itemize}

\subsection{RQ1: Seed reduction by \approach}

We randomly chose one test seed for each test subject, and apply \approach to reduce them. We specified the coverage similarity and reduction ratio constraints to 75\% and 40\% respectively (i.e. the reduced seed must cover at least 75\% of statements covered by the original, and must be at least 40\% smaller than the original). Due to the fact that \approach's core \deltadebugging{} could take long to reduce some seeds that satisfy the conditions, we also allotted a time budget of five minutes to reduce a seed, and rejected seeds not reduceable in five minutes. The seed reduction results are shown in Table~\ref{tab:test-reductions}. In the reduction process, a chunk size of 1 character was used for the \texttt{xmllint} xml seed. For the \texttt{readelf} binary seed, we used a chunk size of 1024 bytes; smaller chunk sizes (16, 256, 512) increased the computation time to beyond 5 mins. Based on these findings we thus observe the following:


\begin{table}[t]
\caption{Reduction results for each seed using \approach ($t_o$ and $t_r$ refer to the original and reduced seed, respectively).}
\centering
\setlength{\tabcolsep}{3pt}
\begin{tabular}{ccccccc}
\hline
\textbf{Seed}      & \textbf{Test}     & \textbf{$t_o$ size}          & \textbf{$t_r$ size} &   \textbf{Size}       & \textbf{Coverage}  & \textbf{Reduction}  \\
\textbf{type}           &  \textbf{Target}   & \textbf{(Bytes)}           & \textbf{(Bytes)} &  \textbf{reduction} & \textbf{similarity} &\textbf{time}  \\ \hline
text (xml)     & \texttt{xmllint}     & 119                  & 19           & 84.03\%        & $>$75\%               & $<$5 mins              \\
binary         & \texttt{readelf}     & 133,432              & 1024       &  99.23\%        & $>$75\%             & $<$5 mins               \\
 \hline          
\end{tabular}

\label{tab:test-reductions}
\end{table}

\noindent \textbf{Observation 1.} \textit{\approach can significantly reduce a seed (by more than 80\%) in less than 5 minutes; the reduced seed it generates can cover more than ${3/4}^{th}$of the statements covered by the original seed.}

\subsection{RQ2: Reduced Seeds and New Path Exploration}

In order to investigate the impact of a reduced seed on AFL's fuzzing performance, we fuzzed each test subject three times with an original (un-reduced) seed and three times with a corresponding reduced seed, for a duration of 24 hours. We performed three repetitions for each fuzz job to capture output variations that arise from the random nature of AFL. The results are illustrated in Figure~\ref{fig-path-results}.

In fuzzing \texttt{readelf}, we observed that AFL finds new paths much faster on average when starting with the reduced seed than it does with the original seed -- a trend seen throughout the 24 hour fuzzing period. On \texttt{xmllint}, a faster pace of path finding was seen with the reduced seed for the first five hours, after which it slowed down. The slow-down was also seen with the original seed but after about 6.5 hours. Another interesting trend seen in both targets is the fuzzer's tendency to stagnate especially with the original seed. For e.g., on \texttt{readelf}, the fuzzer stopped making any progress for several long durations with the original seed. On the other hand, the reduced seed contributed to almost no stagnation, and thus found significantly more paths. For \texttt{xmllint}, although using the original seed resulted in slightly more paths being found over the session, a similar trend was observed where the reduced seed fuzz trajectory broke free from stagnation more often -- leaving it to be seen whether a longer fuzzing duration might have helped its progress. On average, AFL found 1,569 and 6,951 number of paths in fuzzing \texttt{readelf} with the original seed and the reduced seed respectively. In the same order, it found 3,658 and 3,478 number of paths in fuzzing \texttt{xmllint}. We make the following observation:


\begin{figure}[t]
  \centering
   \includegraphics[width=\columnwidth]{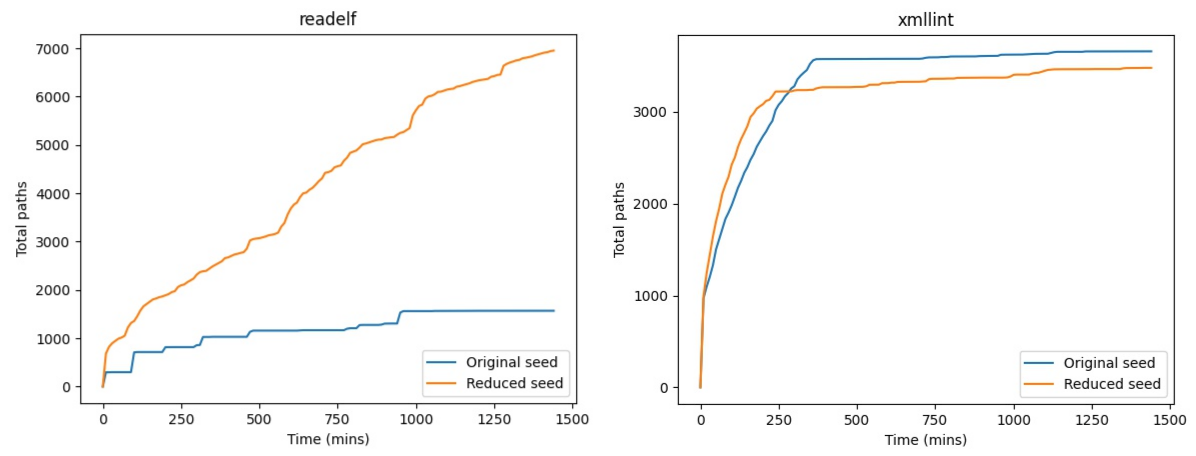}
  \caption{Number of paths found by AFL with the original and the reduced seeds as input, in a 24-hour fuzzing session, on \texttt{readelf} and \texttt{xmllint}. (All numbers are based on the average number of paths found over three fuzzing instances.)}
  \label{fig-path-results}
\end{figure}

\begin{table}[t]
\caption{Number of crashes found with original seed and reduced seed, during each  fuzzing instance.}
\centering
\setlength{\tabcolsep}{10pt}
\begin{tabular}{c|ccc|ccc}
\hline
\multirow{2}{*}{\textbf{Target}}    & \multicolumn{3}{|c}{\textbf{Original Seed}} & \multicolumn{3}{|c}{\textbf{Reduced Seed}} \\ 
    & \textit{job-1}   & \textit{job-2} & \textit{job-3}  & \textit{job-1} & \textit{job-2} & \textit{job-3}    \\ \hline
\texttt{xmllint}     &   0    &   0    &   0    &   0    &   0    &   0       \\
\texttt{readelf}      &   1    &   0    &   0    &   95   &  65    &   106        \\
 \hline          
\end{tabular}

\label{tab:test-reductions:crash}
\end{table}

\noindent \textbf{Observation 2.} \textit{
AFL found significantly more new paths with the \approach-reduced seed than with the original seed while fuzzing \texttt{readelf} (more than four times as many). 
}

\subsection{RQ3: Impact of Reduced Seeds on the Number of Crashes and Coverage}

Table~\ref{tab:test-reductions:crash} depicts the number of crashes in each instance of the fuzzing campaign, and Table~\ref{tab:test-reductions:covarage} depicts the total coverage in the fuzzing campaigns. 
Fuzzing campaigns based on the reduced seed in \texttt{readelf} induced a lot more crashes (266 crashes in total) than campaigns with the original seed (only one crash in total). 
For coverage, in \texttt{xmllint}, fuzzing campaigns that used the original seed, provided higher line and branch coverage in total than  campaigns that used the reduced seed. 
However, in \texttt{readelf} it is the reverse. That is, fuzzing campaigns that used the reduced seed, provided higher line and branch coverage in total than the campaigns that used the original seed.

\noindent \textbf{Observation 3.} \textit{In \texttt{readelf}, fuzzing campaigns with the reduced seeds could cause significantly more crashes and more coverage than the campaigns that used original seeds. 
In \texttt{xmllint}, fuzzing campaigns with the original seeds covered more statements and branches than the campaigns that used reduced seeds.
}

\begin{table}[t]
\caption{Number of lines ($L$) and number of branches ($B$) covered by all seeds generated by AFL from the original seed and the reduced seed.}
\centering
\setlength{\tabcolsep}{18pt}
\begin{tabular}{c|cc|cc}
\hline
\multirow{2}{*}{\textbf{Target}}    & \multicolumn{2}{|c}{\textbf{Original Seed}} & \multicolumn{2}{|c}{\textbf{Reduced Seed}} \\ 
    & $L$  & $B$ & $L$  & $B$   \\ \hline
\texttt{xmllint}     &   6,211    &   7,198    &   5,650    &   6,583          \\
\texttt{readelf}      &   3,004   &   2,882    &   4,472    &   3,836        \\
 \hline          
\end{tabular}
\label{tab:test-reductions:covarage}
\end{table}

\section{Discussion and Future Work}
\label{sec-disc}

Fuzzing can be seen as a search process in the state space of programs. A fuzzer starts with initial seeds, and gradually explores the proximity of the seeds. 
The search objective of fuzzing is to find and exercise interesting behaviors in a program. While the search strategy, including mutation operators and mutation scheduling algorithms, can impact the efficiency in reaching the search objectives, the starting point, i.e. the original seeds, are also important. 

We observed conflicting results between \texttt{readelf} and \texttt{xmllint}. While \approach had positive impact on fuzzing the former, it had a slightly negative impact on the latter. We note that \approach removes over $120$K bytes in the \texttt{readelf} seed but retains $1024$ bytes at the end, while \approach removes $100$ bytes in \texttt{xmllint} and retains merely $19$ bytes. $19$ bytes can limit the fuzzer to efficiently explore new behaviors of the program. We speculate that the non-adequate reduction in \approach removes some useful bytes in the original seed key in covering new paths, and given the time budget AFL could not compensate for that. Perhaps for seeds that are already small, like in \texttt{xmllint}, we should use the original seed, or we use adequate test reduction~\cite{icst14,STVR:STVR1574}.

Having many uninteresting bytes in the seeds impedes effective fuzzing. Therefore, techniques that identify and remove these bytes can positively impact the fuzzing. \approach is our preliminary solution to this problem. The parameters chosen in the experiments, e.g. coverage similarity 75\% etc., are somewhat arbitrary and further investigations are required to systematically assign those parameters given a seed, the software under test, and a fuzzer. 

Another issue in the non-adequate reduction of seeds is that coverage of some parts of the program is lost, and we do not know if the fuzzer can easily cover those parts starting from the reduced seeds. For example, if the lost coverage belongs to a hard to cover part of the program, the fuzzer needs to spend a significant amount of time to cover that part of the program. In such cases, perhaps some analysis of the covered statements/branches is needed to inform the test reduction about the cost of losing coverage of certain parts of code and potentially avoiding that. 

Finally, we note that fuzzing is a stochastic process and different techniques can show different behavior with different fuzzers, seeds, and programs under test. Therefore, the approach we suggested in this paper, may not generalize to all settings as we observed in the \texttt{xmllint} example. 
A potential future work can be to characterize the circumstances under which techniques similar to \approach tend to improve the performance of fuzzing campaigns. 
\section{Related Work}
\label{rel-work}

In this section, we first discuss closely associated work to DIAR's \deltadebugging{} adaptation for test case reduction. Then, since our work aims to analyze the impact of test case reduction on AFL's performance, we present some of the techniques that have been integrated in coverage guided fuzzers to improve their performance.


\noindent\textbf{\Deltadebugging{} based test case reduction}. The traditional \deltadebugging{} algorithm~\cite{zeller-tse-2002} accepts a failing \testcase{} as input and iteratively reduces it by testing candidate sub-parts and removing parts that do not impact the failure outcome, until the smallest such test case is obtained. This work was extended to perform reduction while preserving arbitrary properties of the input test, rather than just its failing property~\cite{icst14,STVR:STVR1574}. To speed up computation of reduced seeds, the property preserving requirement in \deltadebugging{} was relaxed in Alipour et al.'s reduction approach~\cite{non-adequate-reduction} in order to allow candidate test cases to be non-adequate, i.e. to \textit{partially} preserve some property. The fact non-adequate test cases can also be useful, further emphasized the practicability of this modified \deltadebugging{} method, which was thereby adapted in DIAR. Analogously, this modified approach was also used in evaluating neural code intelligence models, where it was adapted to reduce input programs~\cite{prog-simp-ci-model}. Other optimizations to \deltadebugging{} include hierarchical \deltadebugging{}~\cite{hdd}, where test cases are reduced in a structure-aware manner in order to minimize the generation of invalid tests, and parallelization~\cite{Renta2019FuzzTA}.

\noindent\textbf{Approaches for improving coverage guided fuzzing.} While there have been many approaches to improve fuzzing performance, here we discuss four strategies that have been deployed: 

\noindent\textit{1. Internal test case reduction}. In order to introduce smaller test case generation within fuzzing, Hatfield-Dodds developed Hypofuzz~\cite{hypofuzz}, a fuzzing backend that leverages Hypothesis~\cite{hreducer}, a popular property-based testing library for Python. Hypothesis uses \textit{internal reduction}~\cite{hreducer-insights}, which manipulates the test case generator such that the generator itself outputs smaller interesting test cases. It does so by manipulating the random behavior of pseudo-random number generators (PRNG), which are typically used in test case generators. A PRNG generates choice sequences of bits, 
where each bit corresponds to a non-deterministic binary decision. By using shortlex 
optimization~\cite{hreducer-insights}, Hypothesis reduces the lengths of these choice sequences 
in order to generate smaller test cases. 

\noindent\textit{2. Selective tracing}. Wasteful tracing in fuzzers has been targetted by researchers. E.g. UnTracer~\cite{nagy-sp19} improves the efficiency of coverage-guided fuzzers like AFL and Driller~\cite{afl,driller} by eliminating tracing of generated test cases that do not increase coverage\footnote{In their evaluation, they found less than 1 in 10,000 of all test cases are coverage-increasing in conventional coverage guided fuzzers.}. They instrument the
target binary such that the binary self-reports whether a \testcase{} increased
its coverage, thereby gaining in efficiency from the native speeds of the
binaries' executions. 

\noindent\textit{3. Structure-aware fuzzing}. Wasteful fuzzer performance is also contributed by the generation of meaningless test cases, which has been addressed by grammar-based fuzzers like Peach~\cite{peach}, LangFuzz~\cite{langfuzz}, Sulley~\cite{sulley}, CSmith~\cite{csmith}, and Jsfunfuzz~\cite{jsfunfuzz}. However, the need to know and provide grammar specifications has been a major drawback of grammar-based fuzzers.  
Grimoire~\cite{blazytko-sec19} overcomes this by learning the structure of the input during fuzzing, handling inputs in libxml2,
SQLite, Lua, and JavaScriptCore. A similar tool that also learns from inputs is 
Skyfire~\cite{skyfire}. 

\noindent \textit{4. Smart Mutation Scheduling}. Coverage-guided fuzzers have also been made more efficient by modifying their testcase mutation scheduler (for e.g.~\cite{mopt,fuzzergym,reinf-fuzz}). MOPT~\cite{mopt} has been shown to be among the most effective of these techniques in finding bugs -- it uses a modified version 
of the particle swarm optimization algorithm to find an optimal selection probability 
distribution of mutation operators.

\section{Conclusion}
\label{concl}
In this paper, we presented \approach, a technique for preprocessing test seeds for fuzzers. \approach reduces the size of the seeds based on the coverage of the original seeds to remove uninteresting bytes in the seed that can potentially degrade the performance of fuzzing. 
The initial results are promising and warrant further research on characterizing the circumstances in which \approach is applicable.

\clearpage

%
%

%
%
%
\bibliographystyle{splncs04}
\bibliography{ref,ref2,ref3_ase16}
\newpage



\end{document}